\pdfoutput=1
\documentclass[11pt,twoside,a4paper,cmspaper,final,collab]{cms-tdr}
\def\svnVersion{a47739e}\def\svnDate{2022/04/15}\def\cmsCernNoTag{CERN-EP-2021-141}\def\cmsCernDate{\today}\def\cmsMessage{Published in Physics Letters B as \href{http://dx.doi.org/10.1016/j.physletb.2022.137062}{\doi{10.1016/j.physletb.2022.137062}.}}
\begin{document}\cmsNoteHeader{HIN-19-011}

\newlength\cmsFigWidth

\ifthenelse{\boolean{cms@external}}{\setlength\cmsFigWidth{0.98\columnwidth}}{\setlength\cmsFigWidth{0.69\textwidth}}
\ifthenelse{\boolean{cms@external}}{\newcommand{\cmsLeft}{upper\xspace}}{\newcommand{\cmsLeft}{left\xspace}}
\ifthenelse{\boolean{cms@external}}{\newcommand{\cmsRight}{lower\xspace}}{\newcommand{\cmsRight}{right\xspace}}

\newcommand{\sqrts}{ \ensuremath{\sqrt{s}}\xspace}
\newcommand{\RAA}{\ensuremath{R_{\mathrm{AA}}}\xspace}
\newcommand{\TAA}{\ensuremath{T_{\mathrm{AA}}}\xspace}
\newcommand{\PbPb}{\ensuremath{\mathrm{PbPb}}\xspace}
\newcommand{\pp}{\Pp{}\Pp}
\newcommand{\pPb}{\Pp{}\ensuremath{\mathrm{Pb}}\xspace}
\newlength\cmsTabSkip\setlength{\cmsTabSkip}{1ex}
\newcommand{\Pphi}{\PGfP{1020}}
\newcommand{\Bzerosdecayall}{\ensuremath{\PBzs \to \PJGy\,\Pphi \to \Pgmp\Pgmm \PKp \PKm}\xspace}
\newcommand{\Bplusdecayall}{\ensuremath{\PBp \to \PJGy\,\PKp \to \Pgmp\Pgmm \PKp}\xspace}
\newcommand{\Bzerosdecayshort}{\ensuremath{\PBzs \to \PJGy (\Pgmp\Pgmm)\,\phi (\PKp \PKm)}\xspace}
\newcommand{\Bplusdecayshort}{\ensuremath{\PBp \to \PJGy (\Pgmp\Pgmm)\,\PKp}\xspace}
\newcommand{\Bzerosdecay}{\ensuremath{\PBzs \to \PJGy\,\Pphi}\xspace}
\newcommand{\Bplusdecay}{\ensuremath{\PBp \to \PJGy\,\PKp}\xspace}
\newcommand{\Jpsidecay}{\ensuremath{\PJGy\to\PGmp\PGmm}\xspace}
\newcommand{\dstardecay}{\ensuremath{\PDst \to \PK \PGp\PGp(\PGp\PGp)\xspace}}
\newcommand{\phidecay}{\ensuremath{\Pphi\to \PKp \PKm \xspace}}
\newcommand{\npart}{\ensuremath{\langle N_{\text{part}}\rangle}\xspace}
\newcommand{\NMB}{\ensuremath{N_{\text{MB}}}\xspace}
\newcommand{\acceff}{\ensuremath{\alpha\,\epsilon}\xspace}
\newcommand{\acceffinv}{\ensuremath{1/(\acceff)}\xspace}
\newcommand{\meanacceffinv}{\ensuremath{\langle\acceffinv\rangle}\xspace}

\cmsNoteHeader{HIN-19-011}

\title{Observation of \texorpdfstring{\PBzs}{B0s} mesons and measurement of the \texorpdfstring{\PBzs/\PBp}{B0s/B+} yield ratio in \texorpdfstring{\PbPb}{PbPb} collisions at \texorpdfstring{$\sqrtsNN = 5.02\TeV$}{sqrt(sNN) = 5.02 TeV}}

\author{The CMS Collaboration}

\date{\today}

\abstract{
The \PBzs  and \PBp  production yields are measured in PbPb collisions at a center-of-mass energy per nucleon pair of 5.02\TeV. The data sample, collected with the CMS detector at the LHC, corresponds to an integrated luminosity of 1.7\nbinv. 
The mesons are reconstructed in the exclusive decay channels \Bzerosdecayshort and \Bplusdecayshort, in the transverse momentum range 7--50\GeVc and absolute rapidity 0--2.4. The \PBzs meson is observed with a statistical significance in excess of five standard deviations for the first time in nucleus-nucleus collisions.
The measurements are performed as functions of the transverse momentum of the B mesons and of the PbPb collision centrality. The ratio of production yields of \PBzs and \PBp is measured and compared to theoretical models that include quark recombination effects.
}

\hypersetup{
pdfauthor={CMS Collaboration},
pdftitle={Observation of Bs mesons and measurement of the Bs/B+ yield ratio in PbPb collisions at sqrt(s[NN])=5.02 TeV},
pdfsubject={CMS},
pdfkeywords={CMS, quark gluon plasma, heavy flavor, B mesons, hadronization}}

\maketitle

\section{Introduction}

Relativistic heavy ion collisions allow the study of quantum chromodynamics (QCD) at high energy densities and temperatures. Under such conditions, a state in which quarks and gluons are the relevant degrees of freedom, the quark-gluon plasma (QGP)~\cite{QGP1,QGP2}, is formed~\cite{Busza:2018rrf} as predicted by lattice QCD calculations~\cite{Karsch:2003jg}.
Multiple probes are necessary for characterizing the properties of the QGP medium.
Among these, heavy quarks, which are abundantly produced at the CERN LHC, have the potential of providing novel insights into QCD calculations, serving as probes of the QGP~\cite{Andronic:2015wma,Dong:2019byy}.
As they traverse the QGP, these hard-scattered partons lose energy by means of elastic collisions and medium-induced gluon radiation~\cite{Eloss1,Baier:2000mf,Chatrchyan:2011sx,Aad:2010bu}.
The study of parton energy loss can provide insights into the energy density and diffusion properties of the QGP.
The full reconstruction of beauty and charm gives access to their four-momenta and allows the study of the flavor and mass dependences of such processes.
Although heavy flavor data became available at BNL RHIC and at the LHC~\cite{Andronic:2015wma}, there are still large theoretical uncertainties on the hadronization of heavy quarks in the presence of the QGP~\cite{Rapp:2018qla}.
To extract the transport properties of the QGP using heavy quarks, an understanding of the hadronization mechanism is necessary.

An enhancement in the strangeness content of the medium formed in heavy ion collisions~\cite{Abelev:2013xaa,Agakishiev:2011ar,Abelev:2008zk,Arsene:2009jg,Adamczyk:2017wsl,Adare:2015ema,ALICE:2017jyt} is expected when the medium temperature lies above the strange quark mass ~\cite{PhysRevLett.48.1066}.
Through a quark-recombination mechanism~\cite{Molnar:2003ff,Greco:2003mm,Greco:2003vf,He:2014cla}, a corresponding yield enhancement of hadrons containing strangeness is expected relative to corresponding hadrons that do not contain strange quarks. 
Hints of the presence of such recombination processes for heavy mesons containing strangeness were recently reported for open-charm~\cite{Acharya:2018hre} and open-beauty~\cite{BsPbPbCMS} mesons in lead-lead (\PbPb) collisions at a nucleon-nucleon center-of-mass energy of $\sqrtsNN=5.02\TeV$ at the LHC.  

The production of \PB mesons employing fully reconstructed decays was studied at the LHC in proton-proton (\pp) collisions at center-of-mass energies, $\sqrts$, of 7, 8 and 13\TeV~\cite{CMSBmesonpp,Chatrchyan:2011pw,Chatrchyan:2011vh,ATLAS:2013cia,LHCb:2012sng,LHCb:2013JHEP,LHCb:2011leg,Aaij:2012dd,Khachatryan:2014nfa,Aaij:2015fea,Aaij:2014ija,ATLAS:2019jpi,Khachatryan:2016csy,LHCb:2017vec,LHCb:2019tea} over wide transverse momentum (\pt) and rapidity ($y$) intervals, and in proton-lead ($\pPb$) collisions at $\sqrtsNN$ of 5.02 and 8.16\TeV~\cite{Khachatryan:2015uja,LHCb:2019avm}. In \PbPb collisions, the CMS Collaboration reported results at $\sqrtsNN = 5.02$\TeV for the \PBp~\cite{BpPbPbCMS} and \PBzs~\cite{BsPbPbCMS} mesons.
In this Letter, we extend these results by employing the \PbPb data set collected at the end of 2018, a sample larger by about a factor of three as compared to the first measurement with 2015 data, resulting in increased statistical precision and significance.  

The \PBp ($\PQu\PAQb$) and \PBzs ($\PQs\PAQb$) mesons
are reconstructed via the exclusive decay channels \Bzerosdecay and \Bplusdecay, with \Jpsidecay and \phidecay.
Throughout this Letter, unless otherwise specified, the $y$ and \pt variables given are those of the \PB mesons, and charge-conjugate states are implied.
The production yields of the \PBzs and \PBp mesons in \PbPb collisions at $\sqrtsNN = 5.02$\TeV and their ratios are reported, using a data set corresponding to an integrated luminosity of 1.7\nbinv collected by the CMS experiment.
The measurements are performed as functions of the meson \pt and of the \PbPb collision centrality,
where the latter characterizes the degree of overlap of the two lead nuclei and is related to the temperature of the created medium.

\section{Experimental apparatus and data sample}

The central feature of the CMS detector is a superconducting solenoid, which provides a magnetic field of 3.8\unit{T}. Within the solenoid volume are a silicon tracker that measures charged particles in the pseudorapidity range $\abs{\eta} < 2.5$, a lead tungstate crystal electromagnetic calorimeter, and a brass and scintillator hadron calorimeter. For charged particles of $1 < \pt < 10\GeVc$ and $\abs{\eta} < 1.4$, the track resolutions are typically 1.5\% in \pt and 25--90 (45--150)\mum in the transverse (longitudinal) impact parameter \cite{TRK-11-001}. Muons are measured in the range $\abs{\eta} < 2.4$, with detection planes made using three technologies: drift tubes, cathode strip chambers, and resistive-plate chambers. The muon reconstruction algorithm starts by finding tracks in the muon detectors, which are then fitted together with tracks reconstructed in the silicon tracker to form \emph{global muons}. Matching muons to tracks measured in the silicon tracker results in a relative \pt resolution for muons with $20 < \pt < 100\GeVc$ of 1.3--2.0\% in the barrel and better than 6\% in the endcaps.
The hadron forward (HF) calorimeter uses steel as an absorber and quartz fibers as the sensitive material. The two halves of the HF are located 11.2\unit{m} away from the interaction point, one on each end, providing together coverage in the range $3.0 < \abs{\eta} < 5.2$. In this analysis, the HF information is used to select \PbPb collision events and to define their centrality class. Centrality, defined as the fraction of the total inelastic hadronic cross section with 0\% representing collisions with the largest overlap of the two nuclei, is determined experimentally using the total energy in both HF calorimeters~\cite{centralityref}. 
Events are filtered using a two-tiered trigger system~\cite{Khachatryan:2016bia}. 
The first level (L1),
composed of custom hardware processors, uses information from the calorimeters and muon detectors to select events within a fixed time interval of less than 4\mus. The second level, known as the high-level trigger (HLT),
consists of a farm of processors running a version of the full event reconstruction software optimized for fast processing.
A detailed description of the CMS experiment and coordinate system can be found in Ref.~\cite{bib_CMS}.

Several samples of Monte Carlo (MC) simulated events are used to evaluate background components, signal efficiencies and detector acceptance corrections. The simulations include signal samples containing only the \PB meson decay channels being measured, and background samples of other \PQb hadron decays also involving \PJGy mesons in the final state. 
Simulated proton-proton collisions are generated with \PYTHIA~8 v212~\cite{Sjostrand:2014zea} tune CUETP8M1~\cite{Khachatryan:2015pea} and propagated through the CMS detector model using the \GEANTfour package~\cite{Allison:2016lfl}. The decay of the \PB mesons is modeled with \EVTGEN 1.3.0~\cite{Lange:2001uf}, and final-state photon radiation is simulated with \PHOTOS 2.0~\cite{Barberio:1990ms}. Each {\PYTHIA}8 event is embedded into a simulated \PbPb collision event generated with \HYDJET~1.8~\cite{Lokhtin:2005px}, which is tuned to reproduce global event properties, such as the charged hadron \pt spectrum and particle multiplicity~\cite{Khachatryan:2016odn}.

The \PbPb collision events were collected with an L1 hardware trigger requiring the presence of two muon candidates, with no explicit momentum threshold, in coincidence with lead bunches crossing at the interaction point. At the HLT one of the muon candidates is reconstructed using information both from the muon detectors and the inner tracker, while only information from the muon detectors is required for the other muon candidate. The two muon candidates were required to have an invariant mass within $1<m_{\Pgm\Pgm}<5\GeVcc$.
For the offline analysis, events have to pass a set of selection criteria designed to reject events from background processes (beam-gas collisions), as described in Ref.~\cite{Khachatryan:2016odn}. Events are required to have at least one reconstructed primary interaction vertex, formed by two or more tracks, with a distance from the center of the nominal interaction region of less than 15\cm along the beam axis. The shapes of the clusters in the pixel detector must be compatible with those expected from particles produced by a \PbPb collision~\cite{Khachatryan:2010xs}. In order to select inelastic hadronic collisions, the \PbPb events are also required to have at least two towers in each of the HF detectors with energy deposits of more than 4\GeV per tower.
Only events with centrality 0--90\% are selected. Collision centrality bins are given in percentage ranges,
with the 0--90\% bin corresponding to the 90\% of the collisions having the largest overlap of the two nuclei. 
The \PbPb sample corresponds to an integrated luminosity of approximately 1.7\nbinv.
This value is indicative only, as the \PbPb yield is normalized by the total number of minimum bias events sampled, $\NMB = 11.8 \times 10^9$. 

\section{Candidate selection and signal extraction}
\label{sec:sel}

Muon candidates are selected within kinematic constraints that ensure that their reconstruction efficiency stays above 10\%. These limits are $\pt^{\Pgm}>3.5\GeVc$ for $\abs{\eta^{\Pgm}}<1.2$, $\pt^{\Pgm}>1.5\GeVc$ for $2.1<\abs{\eta^{\Pgm}}<2.4$, and $\pt^{\Pgm} >  (5.47 - 1.89 \abs{\eta^{\Pgm}}) \GeVc$ in the $1.2<\abs{\eta^{\Pgm}}<2.1$ region. The muons are also required to match the trigger-level muon candidates, and \emph{soft muon} selection criteria are applied to \emph{global muons} (\ie muons reconstructed using the combined information of the tracker and muon detectors), as defined in Ref.~\cite{Chatrchyan:2012xi}.   
Two muons of opposite charge, with an invariant mass within $\pm$150\MeVcc of the world-average \PJGy meson mass~\cite{pdg2018}, are selected to reconstruct a \PJGy candidate, with a mass resolution of typically 15--30\MeVcc, depending on the dimuon rapidity and \pt. The muon pairs are fitted with a common vertex constraint and are kept if the $p$-value of the $\chi^2$ of the fit is greater than 1\%, thus lowering the background from charm and beauty hadron semileptonic decays. 
Similarly, the \Pphi meson candidates are formed with a common vertex constraint between two oppositely charged particle tracks with $\pt>300\MeVc$, both required to pass standard selection criteria described in Ref.~\cite{Khachatryan:2016odn}.
The invariant mass, assuming the world-average charged kaon mass~\cite{pdg2018} for the two tracks, with a resolution of $\sim$3.9\MeVcc, is required to be within 15\MeVcc of the world-average \Pphi meson mass~\cite{pdg2018}.

The \PBzs (\PBp) meson candidates are constructed by combining a \PJGy candidate with a \Pphi (single track) candidate, and requiring that they originate from a common vertex. In the kinematic vertex fit to the dimuon plus two-track (single-track) system, the tracks are assigned the mass of the charged kaon, and the invariant mass of the muon pair is constrained 
to the nominal \PJGy meson mass~\cite{pdg2018}.
The \PB candidate selection is performed via multivariate discriminators, based on a boosted decision tree (BDT) method~\cite{Hocker:2007ht}. The selection is optimized separately for each meson, as well as each individual bin of \pt. 
The discriminating variables employed include:
the $\chi^{2}$ probability of the decay vertex (the probability for the muon tracks from the \PJGy meson decay and the other charged particle track(s) to originate from a common vertex), the decay length (normalized by its uncertainty), the pointing angle (the angle between the line segment connecting the primary and decay vertices and the momentum vector of the \PB meson), the cosine value of the angle between the \PB meson displacement and momentum in the transverse plane, the two-dimensional (2D) distance between the primary and decay vertices of their daughter tracks (normalized by their uncertainties), and the \pt of the daughter charged particle track(s). 
For the \PBzs meson, two additional selection variables are used: 
the distance along the $z$-direction between the primary vertex and the decay 
vertex (normalized by its uncertainty), as well as the absolute difference 
between the reconstructed  \Pphi meson invariant mass and its nominal value.

The BDT training is performed employing simulated \PB signal samples, and background samples taken from data sidebands (candidates with the invariant mass 0.15--0.35\GeVcc away from the \PB meson nominal mass~\cite{pdg2018}).
The signal samples are scaled to the number of \PB candidates predicted by fixed-order plus next-to-leading order logarithmic (FONLL) perturbative QCD calculations ~\cite{FONLLcharmbottomPP1, FONLLcharmbottomPP2, FONLLcharmbottomPP3} for the integrated luminosity of the analyzed data sample.
The background rejection achieved with the BDT selection is larger than 97\%.

\begin{figure}[!ht]
\centering
\includegraphics[width=.49\textwidth]{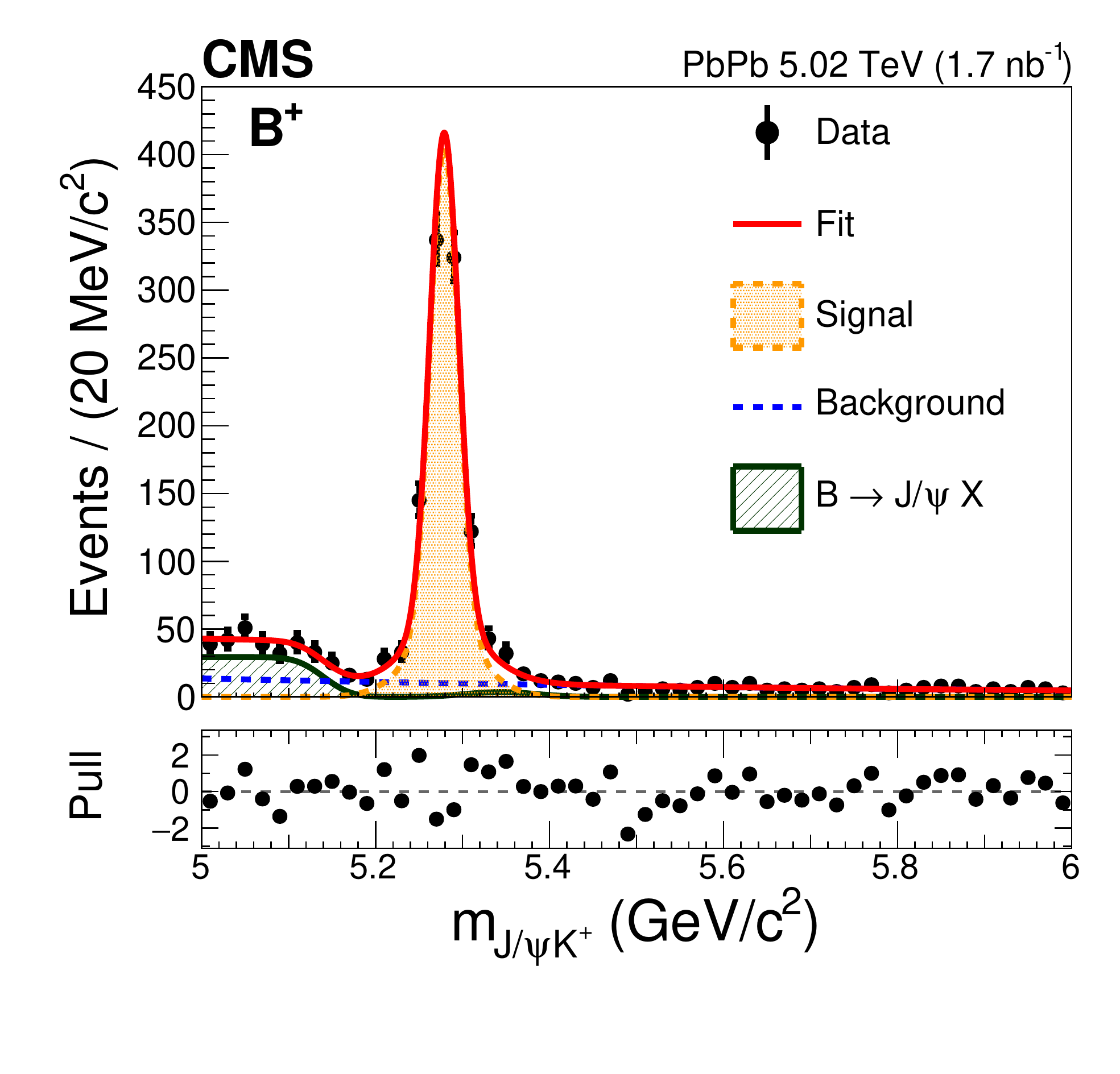}
\includegraphics[width=.49\textwidth]{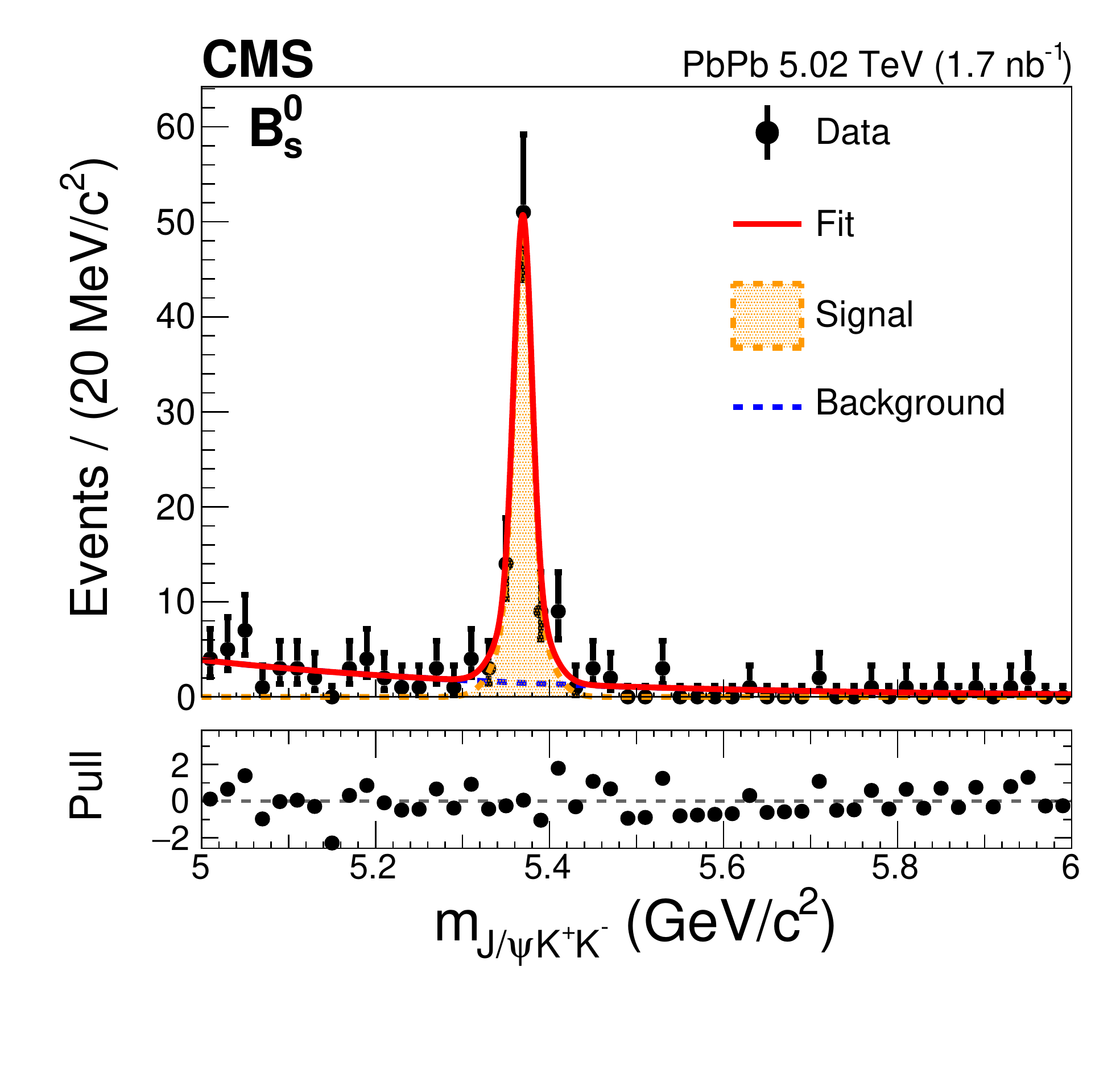}
\caption{Invariant mass distributions of \PBp (\cmsLeft) and \PBzs (\cmsRight) candidates, after the BDT selection,
for event centrality in the range 0--90\%. The lower panels show the pulls, obtained as the difference between the data points and the fit result, divided by the uncertainty in data.}
\label{fig:massPeaks}
\end{figure}

The raw \PB meson signal yields are extracted using an extended unbinned maximum likelihood fit to the invariant mass spectra.
The signal shape is modeled by a sum of two Gaussian functions, with parameters determined from MC simulation, except for the common mean and the overall signal yield, which are free parameters of the fit.
The combinatorial background, from uncorrelated combinations of \PJGy candidates with extra particles,
gives rise to a falling contribution in the invariant mass spectrum that is modeled by an exponential function.
An additional background source can arise from possible contamination from other \PQb hadron decays.
For the \PBp spectrum, partially reconstructed \PB decays, such as $\PBz \to \PJGy\,\PKst^0 \to \Pgmp\Pgmm\PKp\Pgpm$, lead to a heightened background in the invariant mass region below 5.2\GeVcc.
The decay $\PBp \to \PJGy\,\Pgpp$, where the pion is misidentified as a kaon, results in a small peaking structure in the signal region.
Such partially reconstructed and misreconstructed \PQb hadron components are modeled from simulation, via an error function and a Gaussian function, respectively; both shapes and the normalization of the Gaussian function (relative to the signal) are fixed in the fit to the data.   
For the \PBzs meson case, such background contributions are found to be negligible, as a consequence of the tight selection on the mass of the \Pphi candidate.
The results of the fits to the invariant mass distributions are shown in Fig.~\ref{fig:massPeaks}.

The statistical significance of the \PB meson signals is estimated from the ratio of likelihoods obtained by fitting the data with the full model and the background-only model.
Using the standard asymptotic approximation for the likelihood~\cite{Cowan:2010js},
the estimate obtained for the \PBzs meson corresponds to approximately 16 standard deviations.
The estimate is obtained with the nominal signal and background model described above, and verified with the model variations presented in Section~\ref{sec:syst}.  
This is the first observation of the \PBzs signal in heavy ion collisions. 

\section{Production yields}

For each B meson, the production yield scaled by the overlap nuclear function, \TAA~\cite{glauber2007}, and the number of minimum bias events, \NMB, is computed in each \pt interval, according to
\begin{equation}
 \frac{1}{\TAA} \frac{{\rd}N}{{\rd}\pt} =  \frac{1}{2 \, \mathcal{B} \, \NMB \, \TAA}
 \frac{N_{\text{obs}}(\pt)}{\Delta\pt} 
\langle \frac{1}{\alpha(\pt, y) \, \epsilon(\pt, y)} \rangle,
 \label{eq:xsection}
\end{equation}
where $N_{\text{obs}}$ denotes the raw signal yield, extracted in each \pt interval of width $\Delta \pt$. 
The factor 1/2 accounts for the fact that the raw yield is measured for particles and antiparticles added together while the production yield is given for one species only, and $\mathcal{B}$ stands for the branching fraction of the corresponding decay chain.
The factor \TAA is equal to the number of nucleon-nucleon (NN) binary collisions divided by the NN total inelastic cross section. The value of \TAA  can be interpreted as the NN-equivalent integrated luminosity per heavy ion collision.
The \TAA values for  $\sqrtsNN=5.02\TeV$ \PbPb collisions for the centrality intervals 0--30\%, 30--90\%, and 0--90\% are, respectively, $(15.4\pm0.3)\mbinv$, $(1.7\pm0.1)\mbinv$, and $(6.3\pm0.1)\mbinv$, as estimated from a MC Glauber model~\cite{glauber2007}, implemented in the {\textsc{tglaubermc}} v3.2 software package~\cite{glauber2017}.
The measurement is performed in a similar fashion as a function of the \PbPb collision centrality. The centrality class is represented as the average number of participating nucleons, \npart. This number, determined using the above Glauber model, is highly correlated with the impact parameter of the collision.

The acceptance and the efficiency factors, the last term in Eq.~(\ref{eq:xsection}), are obtained employing 2D fine-grained maps of the acceptance $\alpha (\pt,y)$ and the efficiency $\epsilon(\pt,y)$.
These maps are determined, for each centrality region, from MC-simulated samples of \PB meson signal events, generated within the analysis fiducial region given by:
$\abs{y}<1.5$ and $\pt>10\GeVc; 1.5<\abs{y}<2.4$ and $7<\pt<50\GeVc$.
The acceptance corresponds to the fraction of generated events passing the muon and kaon selection thresholds specified in Section~\ref{sec:sel}, while the efficiency is determined as the fraction of accepted events that pass the full analysis selection criteria, including the trigger.
The maps are used to determine the \acceffinv value for each \PB candidate in data, based on the kinematics $(\pt,y)$ of the candidates, and the corresponding average \meanacceffinv obtained for the candidates within a 160\MeVcc window centered on the \PB meson's world-average mass. 
The efficiency maps derived from simulation are corrected by data/MC scale factors for the muon reconstruction and trigger efficiencies, obtained by applying the \emph{tag-and-probe} method using the \PJGy resonance~\cite{Khachatryan:2010zg}.

\begin{table*}[!ht]
\centering
\topcaption{Summary of systematic uncertainties in the \TAA-scaled yield measurements for \PBp and \PBzs mesons, in four $p_T$ ranges. The measurements are performed in the \PB meson kinematic region $\abs{y}<2.4$ and centrality range 0--90\%. The relative uncertainty values are shown in percentage.}
\label{tab:systPt}
  \begin{tabular}{ l  {c}@{\hspace*{5pt}} cccc {c}@{\hspace*{5pt}}  cccc  }
                                    && \multicolumn{4} {c} {\PBp}   && \multicolumn{4} {c} {\PBzs} \\
    \cline{1-1}\cline{3-6}\cline{8-11}
B meson $p_{T}$ (GeV/c)           && 7--10 & 10--15 & 15--20 & 20--50  && 7--10 & 10--15 & 15--20 & 20--50   \\ [\cmsTabSkip]
\multirow{2}{*}{Muon efficiency}
			                && $+$7.2   & $+$4.3  & $+$3.8 &  $+$3.9  && $+$8.9  & $+$6.0   & $+$3.7  & $+$3.9 \\
                                       && $-$6.3  & $-$3.9 & $-$3.5 &  $-$3.6 && $-$7.5  & $-$5.2  & $-$3.5   &   $-$3.6  \\
    Data/MC agreement                       && 4.2 &15    & 3.0   & 1.7 && 35    & 5.6     & 4.7 &10   \\
    MC sample size                      && 9.1   & 3.2 & 1.9  & 1.4   && 27   & 6.3     & 3.1  &3.2  \\
    Fit modeling                     && 4.5    & 2.7 & 2.8   & 2.6   && 1.2   & 3.8     & 1.8   & 6.4    \\
    Tracking efficiency                    && 5.0 & 5.0     & 5.0    & 5.0   && 10   & 10     & 10 &10  \\
     \TAA   && \multicolumn{4} {c} {2.2\;\,} && \multicolumn{4} {c} {2.2\;\,} \\ 
    \NMB   && \multicolumn{4} {c} {1.3\;\,} && \multicolumn{4} {c} {1.3\;\,} \\ 
    Branching fraction && \multicolumn{4} {c} {2.9\;\,} && \multicolumn{4} {c} {7.5\;\,} \\ [\cmsTabSkip]
\multirow{2}{*}{Total}   
					 && $+$15 &  $+$17 &  $+$8.7 &  $+$ 8.2   &&  $+$47 &  $+$17 &  $+$15 &  $+$18  \\
		                          && $-$14 &   $-$17 &  $-$8.5 &  $-$8.0    && $-$47  &  $-$17& $-$14 & $-$18 \\
\hline

\end{tabular}
\end{table*}

\begin{table*}[!ht]
\centering
\topcaption{Summary of systematic uncertainties in the \TAA-scaled yield measurements for \PBp and \PBzs mesons, in three centrality ranges. The measurements are performed in the \PB meson kinematic region given by $10<\pt<50\GeVc$ and $\abs{y}<2.4$. The relative uncertainty values are shown in percentage.}
\label{tab:syst}
  \begin{tabular}{ l  {c}@{\hspace*{5pt}} ccc {c}@{\hspace*{5pt}}  ccc  }
                                    && \multicolumn{3} {c} {\PBp}   && \multicolumn{3} {c} {\PBzs} \\
    \cline{1-1}\cline{3-5}\cline{7-9}
Centrality class           && 0--30\% & 30--90\% & 0--90\%  && 0--30\% & 30--90\% & 0--90\% \\ [\cmsTabSkip]
\multirow{2}{*}{Muon efficiency}
			              && $+$4.2  & $+$4.1   & $+$4.2     && $+$5.5   & $+$4.6  & $+$5.3 \\
                                      && $-$3.8  & $-$3.8  & $-$3.8     && $-$4.9   & $-$4.2  & $-$4.7 \\
    Data/MC agreement                 && 13    & 8.0     & 12       && 3.1    & 3.7   & 3.2  \\
    MC sample size                    && 3.2   & 2.2     & 2.4      && 6.6    & 2.3  & 4.4  \\
    Fit modeling                      && 2.5   & 2.8     & 2.6      && 2.5    & 3.2   & 2.3  \\
    Tracking efficiency               && 5.0   & 5.0     & 5.0      && 10     & 10    & 10   \\
    \TAA                              && 2.0   & 3.6     & 2.2      && 2.0   & 3.6   & 2.2  \\
      \NMB   && \multicolumn{3} {c} {1.3\;\,} && \multicolumn{3} {c} {1.3\;\,} \\ 
    Branching fraction && \multicolumn{3} {c} {2.9\;\,} && \multicolumn{3} {c} {7.5\;\,} \\ [\cmsTabSkip]
    \multirow{2}{*}{Total}   
					 && $+$16 &  $+$12 &  $+$15 &&  $+$16   & $+$15  & $+$15  \\
		                          && $-$15 &   $-$12 &  $-$15 &&  $-$16    &   $-$15  & $-$15 \\
    \hline
\end{tabular}
\end{table*}

\section{Systematic uncertainties}
\label{sec:syst}

The \TAA-scaled yield measurements are affected by several sources of systematic uncertainties arising from the signal extraction, acceptance and efficiency, $\mathcal{B}$, \NMB, and \TAA determinations.

The uncertainty from the signal and background modeling is evaluated by considering the following fit variations:
(i) using low-order polynomials for describing the combinatorial background, 
(ii) using a sum of three Gaussian functions with a common mean for alternatively describing the signal,
(iii) varying the widths of the signal double-Gaussian function by 10\% (to account for possible mismatches in the mass resolution between simulation and data), and
(iv) fixing the double-Gaussian mean to its MC simulation value. 
The maximum of the signal variations and the maximum of the background variations are added in quadrature as the systematic uncertainty from fit modeling.

The systematic uncertainties associated with the limited size of the MC simulation samples are obtained by recomputing the 2D fine-grained $\alpha(\pt,y) \, \epsilon(\pt,y)$ maps and the corresponding \meanacceffinv averages over the data sample.
The statistical uncertainties from the MC-derived product of acceptance and efficiency are propagated by generating 10000 2D maps (per bin in the differential measurement), where each entry is obtained by randomly sampling 
the \acceff values from Gaussian distributions with widths given by the statistical uncertainty. The resulting systematic uncertainties are obtained as the width of the \meanacceffinv distribution from the alternative 2D maps. 

The efficiency and acceptance determinations, based on MC simulation, are further affected by potential disagreements between data and MC.
The signal MC simulation is validated against data by inspecting the distributions of the variables employed in the selection. The signal distributions are extracted from the data employing the sPlot method~\cite{splot}, using the mass of the \PB meson candidates as discriminating variable, and are also cross-checked with a simple sideband-subtraction method. In particular, the sPlot-derived signal distributions for the BDT score are retrieved from the data, and the corresponding data/MC ratios formed. These ratios are in turn used to re-weight the MC simulation, and the resulting deviation in the \meanacceffinv factors are assigned as systematic uncertainties.
This procedure is employed using the higher-yield channel, namely the \PBp.
For the \PBzs meson, the limited size of the data sample yields results that are compatible between data and simulation, within the statistical uncertainties. As such, the \PBp channel is used as calibration mode for the \PBzs channel, and a procedure identical to that described above is applied to the \PBp sample but using only tracking related variables.
The systematic uncertainty for the \PBzs corresponds to that evaluated for the \PBp and further taking the maximum of the deviation determined from tracking related variables to account for the presence of an extra track in the \PBzs channel.

\begin{figure}[!ht]
\centering
\includegraphics[width=.48\textwidth]{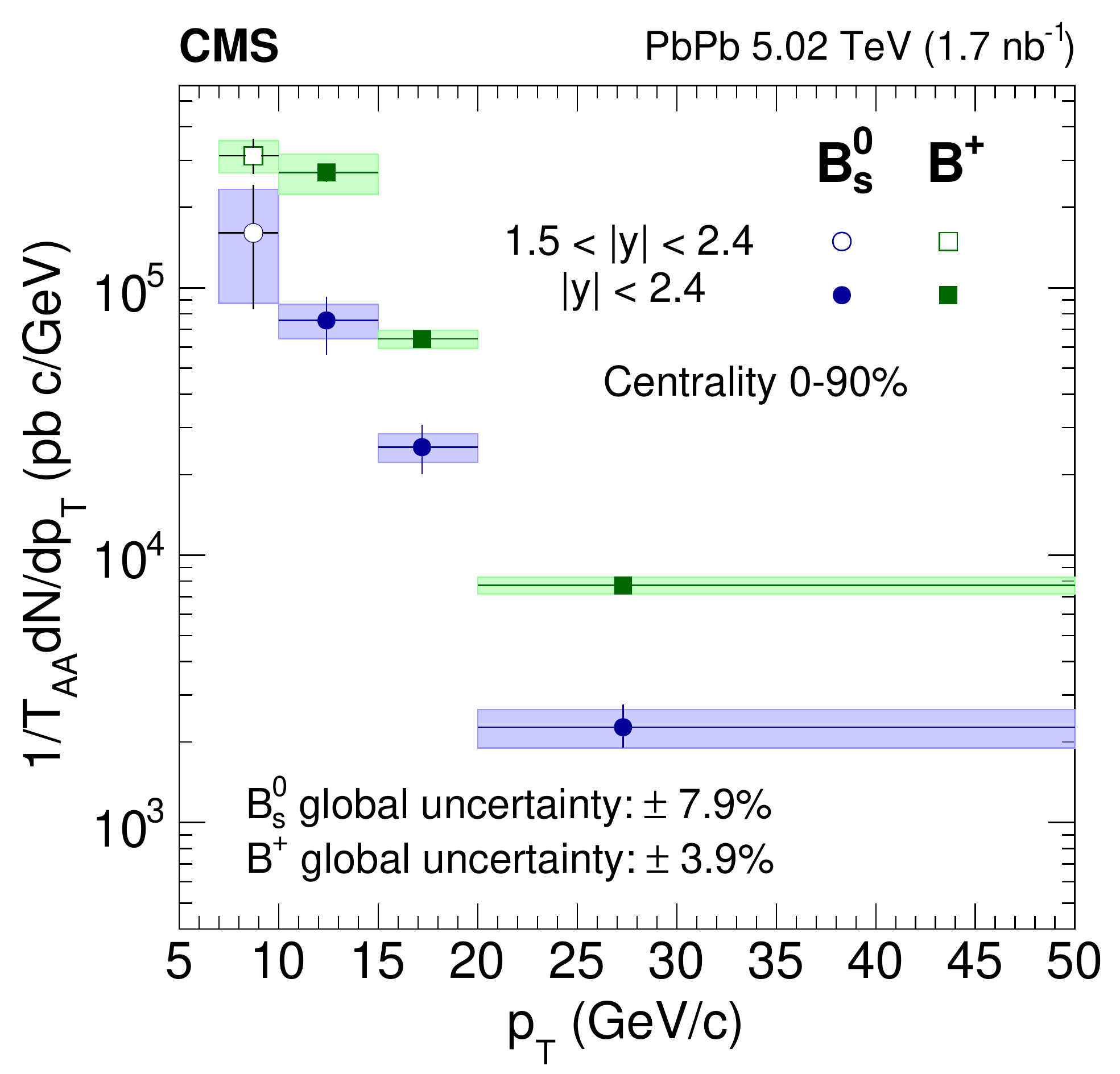}
\includegraphics[width=.48\textwidth]{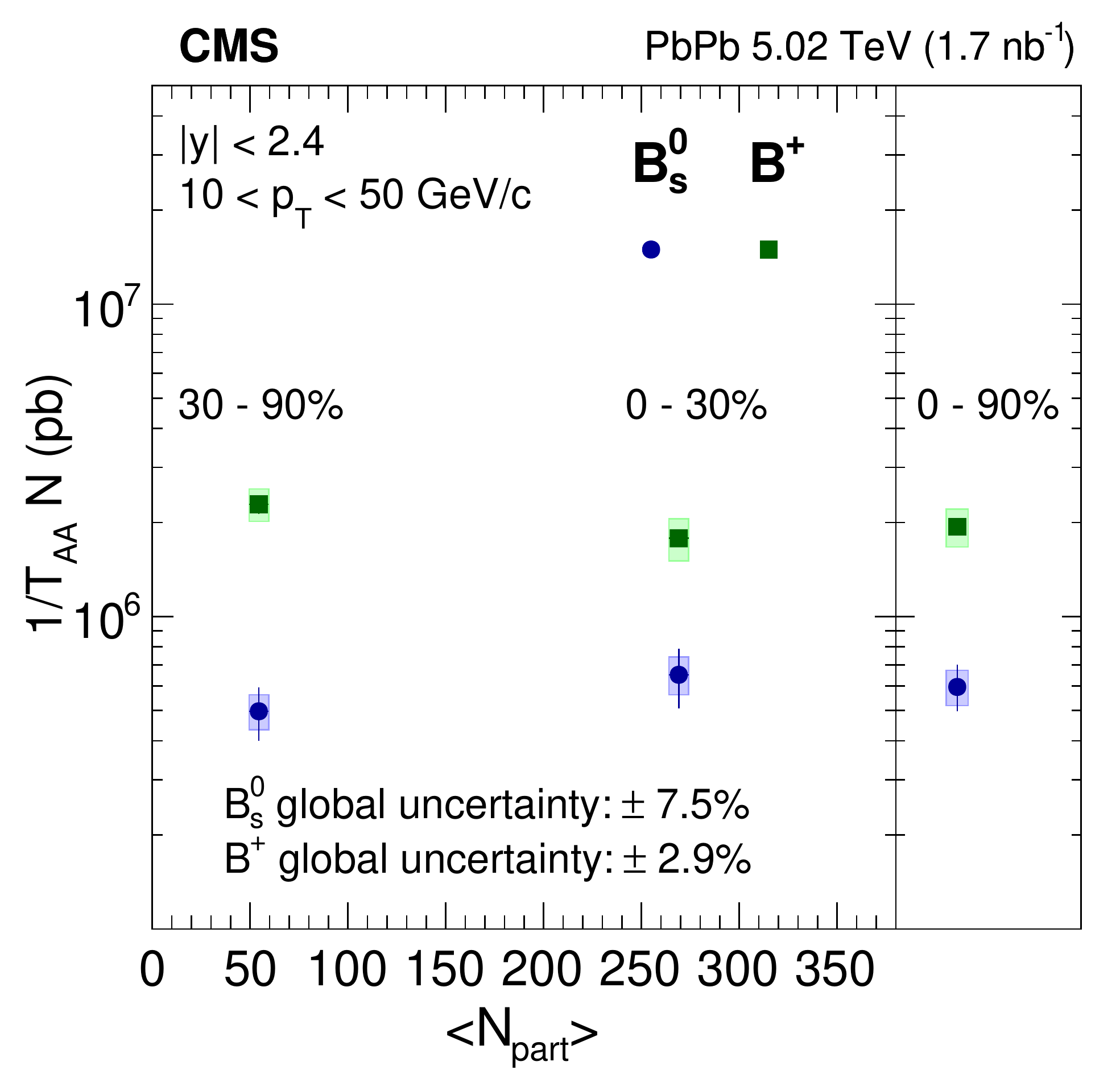}
\caption{The acceptance- and efficiency-corrected yields for the \PBp  and \PBzs mesons, scaled by \TAA and \NMB in \PbPb collisions at $\sqrtsNN=5.02\TeV$. The results are shown as a function of the meson \pt (\cmsLeft), and of the event centrality (\cmsRight), where the rightmost panel indicates the centrality-integrated result.  The vertical bars (boxes) correspond to statistical (systematic) uncertainties. The global systematic uncertainty comprises the uncertainties in \TAA, \NMB, $\mathcal{B}$ (\cmsLeft) and $\mathcal{B}$ (\cmsRight).
}
\label{fig:xsections}
\end{figure}

The uncertainty in the efficiency of the muon trigger, reconstruction, and identification is evaluated using a tag-and-probe technique~\cite{Khachatryan:2010zg}.
The derived data versus MC scale factors are employed in the determination of the nominal \acceff 2D maps, while the associated asymmetric uncertainties are propagated as the systematic uncertainty in the  \meanacceffinv factors.
The difference in the track reconstruction efficiency in data and simulation is estimated by comparing 3-prong and 5-prong \PDst decays, \dstardecay~\cite{Khachatryan:2016odn}. 
 This results in 5\,(10)\% uncertainty in the efficiency determination for the \PBp  (\PBzs) decay for involving one (two) kaon(s).

The determination of the \meanacceffinv factors, for each analysis bin, involves averaging over the data. In order to reflect this statistical effect, the overall statistical uncertainty from the \PbPb data is calculated using a bootstrap method~\cite{bootstrap}, where data events are resampled 1000 times with replacement. The corrected yield is re-determined for each bootstrap sample, and the spread of its distribution is assigned as the statistical uncertainty.

Various consistency checks of the analysis procedure have been performed, which do not result in additional systematic uncertainties.
The consistency of the fitting procedure was verified by generating and fitting pseudo-experiments using the likelihood model and parameters employed for fitting the data. 
The distributions of the pull (here defined as the difference between the generated and fitted parameter values divided by the uncertainty returned by the fit) show random behavior consistent with Gaussians of zero mean and unit width, which indicate the absence of bias in the fitting procedure.
A consistency check was performed using MC simulation, where the signal yields are extracted with the fitting procedure and the acceptance and efficiency corrections applied, thus retrieving the generated yields.
Potential biases in the acceptance and efficiency calculations associated with the shape of the \PB mesons kinematic distributions from simulation were probed by re-weighting the  {\PYTHIA} simulation according to varied \pt shapes
 and recomputing the factors \meanacceffinv. Such deviations were found to be negligible. 
This confirmed independence from the MC simulation is expected in view of the fine-grained 2D \acceff MC maps and the \meanacceffinv averaging based on data, and further attests the robustness of the analysis procedure. 
The effect of background contamination in the averaging procedure was studied employing the sPlot method and was found to be negligible.  

The systematic uncertainties in the Glauber model normalization factor (\TAA) are derived from propagating the uncertainties in the event selection efficiency, and in the nuclear radius, skin depth, and minimum distance between nucleons in the Pb nucleus parameters of the Glauber model~\cite{Khachatryan:2016odn}. The uncertainties associated with the branching fraction of the decay chains $\mathcal{B}$ are obtained from Ref.~\cite{pdg2018}.
The $\mathcal{B}$ factor is common to all bins in the production yield measurements, as is the case also of \TAA and \NMB in the \pt-differential results; the corresponding uncertainties are denoted global uncertainties.  

The total systematic uncertainties in the \TAA-scaled yield measurement are computed as the sum in quadrature of the different contributions mentioned above, which are summarized in Tables~\ref{tab:systPt} and~\ref{tab:syst}.

\begin{figure}[!ht]
\centering
\includegraphics[width=.48\textwidth]{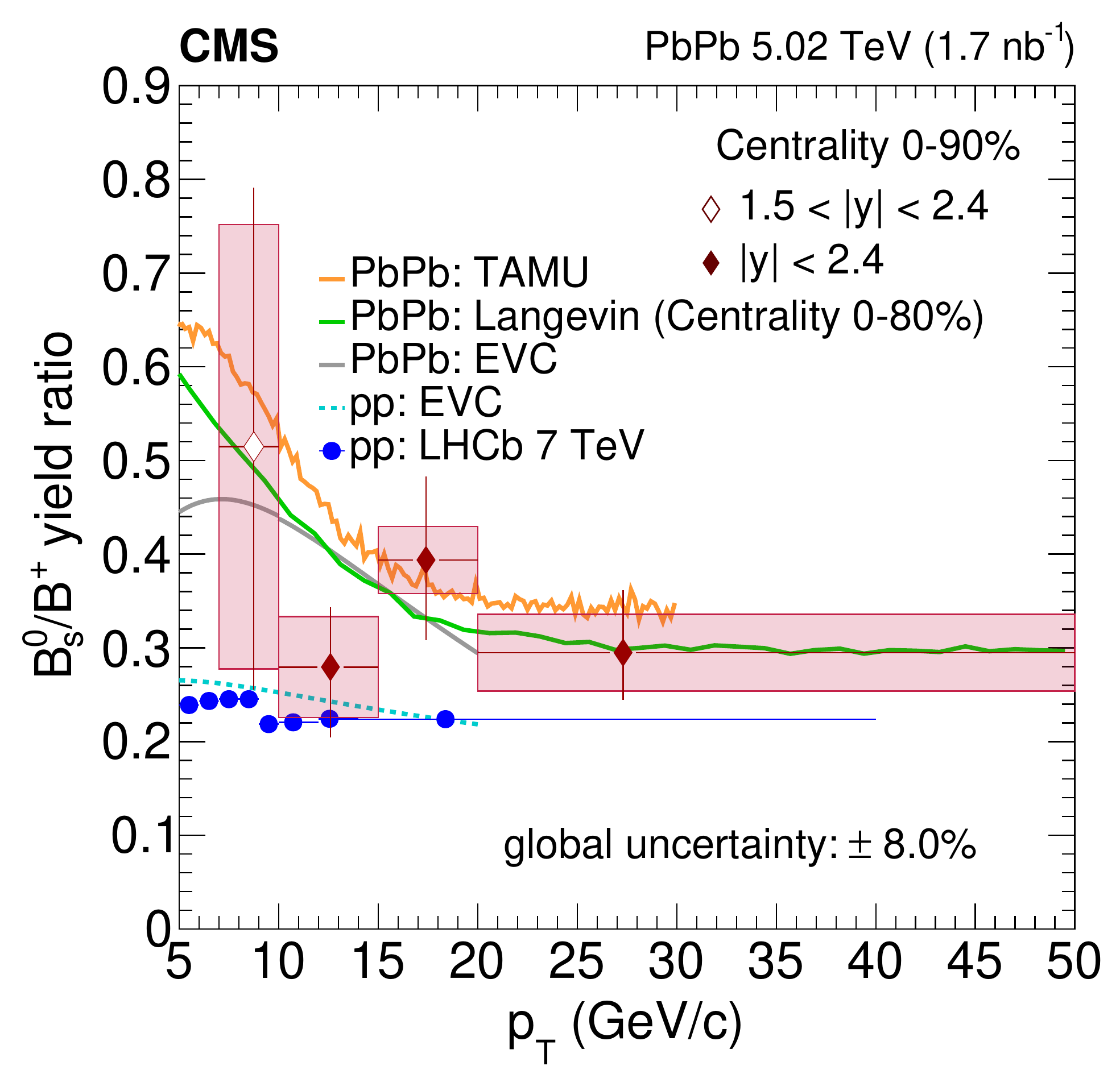}
\includegraphics[width=.48\textwidth]{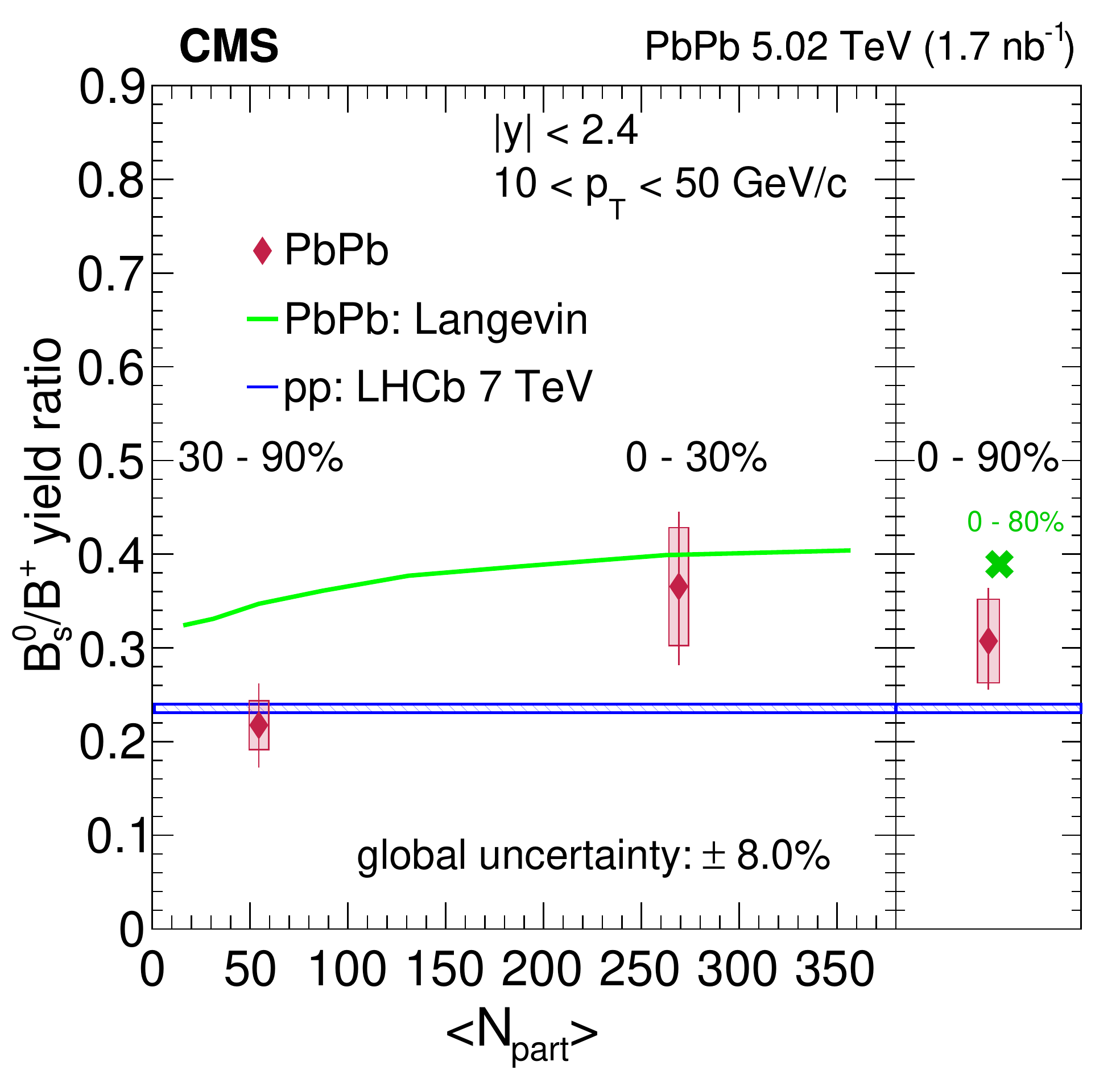}
\caption{The ratio of \PBzs and \PBp meson production yields, as a function of \pt (\cmsLeft) and a function of collision centrality (\cmsRight). The vertical bars (boxes) correspond to statistical (systematic) uncertainties.
The global systematic uncertainty on the yields ratio corresponds to the decay branching fractions $\mathcal{B}$.
Predictions from a Langevin hydrodynamic model~\cite{cao20} and (\cmsLeft panel only) from the TAMU transport model~\cite{He:2014cla} and the EVC quark combination model~\cite{EVC} for \pp (cyan, dashed) and \PbPb (gray) collisions are overlaid.
The yield ratio measured by LHCb in \pp collisions at 7\TeV~\cite{fsfulhcb2020}, scaled by the branching fraction ratio~\cite{pdg2018}, is also displayed (shown in blue, as function of \pt on the \cmsLeft panel, and the average value on the \cmsRight panel, where the double-line width denotes the uncertainty).}
\label{fig:ratio}
\end{figure}

\section{Results}

The differential \TAA-scaled yields for \PBzs and \PBp mesons in \PbPb collisions, Eq.~(\ref{eq:xsection}), 
as a function of the meson \pt are presented in Fig.~\ref{fig:xsections} (\cmsLeft).
On the horizontal scale, the abscissa of each data point is set to the mean value of the \pt distribution, after background subtraction via sPlot, while the bin ranges are represented by the boxes.
In Fig.~\ref{fig:xsections} (\cmsRight), the \TAA-scaled yields are presented for different \PbPb event centrality ranges.

The ratio of production yields of the \PBzs and \PBp mesons is formed as
\begin{equation}
R = \frac{N^{\PBzs}}{N^{\PBp}} = 
\frac{N_{\text{obs}}^{\PBzs}}{N_{\text{obs}}^{\PBp}}
\frac{ \mathcal{B}^{\PBp}}{ \mathcal{B}^{\PBzs}} 
\frac{\langle 1/(\alpha^{\PBzs} \, \epsilon^{\PBzs})\rangle}
     {\langle 1/(\alpha^{\PBp}  \, \epsilon^{\PBp} )\rangle}
.
\label{eq:ratio}
\end{equation}
The terms \TAA and \NMB, that are common to the two meson samples, cancel in the ratio $R$. 
Correlated systematic uncertainties associated with the efficiency determination are partially reduced. 
For the tracking efficiency, a 5\% uncertainty is assigned
because of the different number of hadron tracks in the two decays.
Muon efficiency uncertainties from the tag-and-probe method are determined by varying coherently the efficiency factors in the numerator and denominator within the corresponding uncertainties. 
For the determination of the uncertainties related to data/MC comparisons, an identical procedure to that used for the production yield is employed, but utilizing only tracking related variables, using the \PBp channel. 
The dominant uncertainty arises from data/MC agreement and the limited MC sample size. 
The $R$ ratio results are shown in Fig.~\ref{fig:ratio} 
as functions of the mesons' \pt (\cmsLeft) and event centrality (\cmsRight).

 The \pt dependence of the ratio $R$ is compared to a transport model (TAMU) based on a Langevin equation that includes collisional energy loss and heavy quark diffusion in the medium~\cite{He:2014cla} and an advanced Langevin hydrodynamic model with both heavy quark elastic and inelastic energy loss~\cite{cao20}. Both predictions account for recombination contributions to the \PBzs meson yields, which are expected to be more significant at low \pt. 
Predictions form a quark combination model with the equal-velocity combination (EVC) approximation~\cite{EVC} are also shown for both \PbPb and \pp collisions.
The ratio of the \PBzs and \PBp  yields in \pp collisions has been determined by LHCb~\cite{fsfulhcb2020},   
at different collision energies, for \PB mesons within  $0.5<\pt< 40\GeVc$ and $2<\eta<5$.
While no rapidity dependence of the ratio has been found, a decrease with \pt and an increase with collision energy have been reported~\cite{fsfulhcb2020,LHCb:2021qbv}.
While the $\PBzs$ to $\PBp$ meson yield ratios obtained in charmonium decay channels cannot be directly compared with those measured in $\Pep\Pem$ collisions at LEP~\cite{hflav}, earlier LHCb measurements~\cite{fsfulhcb2013,LHCb:2011leg,fsfulhcb2019} in $\Pp\Pp$ collisions in other hadronic decay channels are compatible with the LEP results. The lack of sizable strangeness enhancement in the $\PBzs$ meson hadronization in $\Pp\Pp$ collisions, evident from the above agreement, makes $\Pp\Pp$ collisions a robust baseline for comparison with our \PbPb results.
In Fig.~\ref{fig:ratio}, the \pp results at 7\TeV by LHCb~\cite{fsfulhcb2020} are scaled by the same ratio of branching fractions~\cite{pdg2018} as used in Eq.~(\ref{eq:ratio}). 
The results are compatible with both the model predictions for \PbPb collisions and the \pp reference results.
However, no significant \pt dependence can be established with the precision allowed by the current data. 
The results, while compatible, tend to lie systematically above the \pp reference data. 
While the central value of $R$ in the more central bin lies above that of the more peripheral bin, the results are compatible within the combined statistical and systematic uncertainty.
These results are compatible with the ratio of nuclear modification factors, \RAA, reported earlier by the CMS Collaboration~\cite{BsPbPbCMS},
albeit with reduced statistical uncertainties.

\section{Summary}

The \PBzs and \PBp mesons are studied with the CMS detector at the LHC via the reconstruction of the exclusive hadronic decay channels \Bzerosdecayall and \Bplusdecayall. 
The measurements are performed within the \PB mesons' fiducial region given by transverse momentum $\pt>10\GeVc$ for rapidity $\abs{y}<1.5$  and  $7<\pt<50\GeVc$ for $1.5<\abs{y}<2.4$.
The first observation of the \PBzs meson in nucleus-nucleus collisions, with a statistical significance well surpassing five standard deviations, is attained. 
The production yields of \PBzs and \PBp mesons, scaled by the nuclear overlap function \TAA and the number of minimum bias events \NMB, in lead-lead (\PbPb) collisions at a center-of-mass energy of $5.02\TeV$ per nucleon pair are presented as functions of the meson \pt and, for the first time, of the event centrality.
These results extend, and are compatible with, those previously reported by the CMS Collaboration~\cite{BsPbPbCMS,BpPbPbCMS}, and are based on a three-fold larger \PbPb data sample.
The ratio of production yields of the two mesons in \PbPb collisions is determined and it is found to be statistically compatible with the corresponding
ratio in proton-proton (\pp) collisions. 
The further investigation of possible hints of an enhancement of the ratio in \PbPb, relative to \pp collisions, will benefit from more precise \PbPb and \pp reference data taken at the same collision energy per nucleon.  
The larger \PbPb data sets that should be accumulated in upcoming high-luminosity LHC heavy ion runs will provide greater precision and could help to further characterize the mechanisms of beauty hadronization in heavy ion collisions.

\begin{acknowledgments}
  We congratulate our colleagues in the CERN accelerator departments for the excellent performance of the LHC and thank the technical and administrative staffs at CERN and at other CMS institutes for their contributions to the success of the CMS effort. In addition, we gratefully acknowledge the computing centres and personnel of the Worldwide LHC Computing Grid and other centres for delivering so effectively the computing infrastructure essential to our analyses. Finally, we acknowledge the enduring support for the construction and operation of the LHC, the CMS detector, and the supporting computing infrastructure provided by the following funding agencies: BMBWF and FWF (Austria); FNRS and FWO (Belgium); CNPq, CAPES, FAPERJ, FAPERGS, and FAPESP (Brazil); MES and BNSF (Bulgaria); CERN; CAS, MOST, and NSFC (China); MINCIENCIAS (Colombia); MSES and CSF (Croatia); RIF (Cyprus); SENESCYT (Ecuador); MoER, ERC PUT and ERDF (Estonia); Academy of Finland, MEC, and HIP (Finland); CEA and CNRS/IN2P3 (France); BMBF, DFG, and HGF (Germany); GSRI (Greece); NKFIA (Hungary); DAE and DST (India); IPM (Iran); SFI (Ireland); INFN (Italy); MSIP and NRF (Republic of Korea); MES (Latvia); LAS (Lithuania); MOE and UM (Malaysia); BUAP, CINVESTAV, CONACYT, LNS, SEP, and UASLP-FAI (Mexico); MOS (Montenegro); MBIE (New Zealand); PAEC (Pakistan); MSHE and NSC (Poland); FCT (Portugal); JINR (Dubna); MON, ROSATOM, RAS, RFBR, and NRC KI (Russia); MESTD (Serbia); SEIDI, CPAN, PCTI, and FEDER (Spain); MoSTR (Sri Lanka); Swiss Funding Agencies (Switzerland); MST (Taipei); ThEPCenter, IPST, STAR, and NSTDA (Thailand); TUBITAK and TAEK (Turkey); NASU (Ukraine); STFC (United Kingdom); DOE and NSF (USA).
   
  \hyphenation{Rachada-pisek} Individuals have received support from the Marie-Curie programme and the European Research Council and Horizon 2020 Grant, contract Nos.\ 675440, 724704, 752730, 758316, 765710, 824093, 884104, and COST Action CA16108 (European Union); the Leventis Foundation; the Alfred P.\ Sloan Foundation; the Alexander von Humboldt Foundation; the Belgian Federal Science Policy Office; the Fonds pour la Formation \`a la Recherche dans l'Industrie et dans l'Agriculture (FRIA-Belgium); the Agentschap voor Innovatie door Wetenschap en Technologie (IWT-Belgium); the F.R.S.-FNRS and FWO (Belgium) under the ``Excellence of Science -- EOS" -- be.h project n.\ 30820817; the Beijing Municipal Science \& Technology Commission, No. Z191100007219010; the Ministry of Education, Youth and Sports (MEYS) of the Czech Republic; the Deutsche Forschungsgemeinschaft (DFG), under Germany's Excellence Strategy -- EXC 2121 ``Quantum Universe" -- 390833306, and under project number 400140256 - GRK2497; the Lend\"ulet (``Momentum") Programme and the J\'anos Bolyai Research Scholarship of the Hungarian Academy of Sciences, the New National Excellence Program \'UNKP, the NKFIA research grants 123842, 123959, 124845, 124850, 125105, 128713, 128786, and 129058 (Hungary); the Council of Science and Industrial Research, India; the Latvian Council of Science; the Ministry of Science and Higher Education and the National Science Center, contracts Opus 2014/15/B/ST2/03998 and 2015/19/B/ST2/02861 (Poland); the Funda\c{c}\~ao para a Ci\^encia e a Tecnologia, grant CEECIND/01334/2018 (Portugal); the National Priorities Research Program by Qatar National Research Fund; the Ministry of Science and Higher Education of the Russian Federation, project no. 14.W03.31.0026 (Russia); 
 the Programa Estatal de Fomento de la Investigaci{\'o}n Cient{\'i}fica y T{\'e}cnica de Excelencia Mar\'{\i}a de Maeztu, grant MDM-2015-0509 and the Programa Severo Ochoa del Principado de Asturias; the Stavros Niarchos Foundation (Greece); the Rachadapisek Sompot Fund for Postdoctoral Fellowship, Chulalongkorn University and the Chulalongkorn Academic into Its 2nd Century Project Advancement Project (Thailand); the Kavli Foundation; the Nvidia Corporation; the SuperMicro Corporation; the Welch Foundation, contract C-1845; and the Weston Havens Foundation (USA).
\end{acknowledgments}

\bibliography{auto_generated}

\cleardoublepage \appendix\section{The CMS Collaboration \label{app:collab}}\begin{sloppypar}\hyphenpenalty=5000\widowpenalty=500\clubpenalty=5000\vskip\cmsinstskip
\textbf{Yerevan Physics Institute, Yerevan, Armenia}\\*[0pt]
A.~Tumasyan
\vskip\cmsinstskip
\textbf{Institut f\"{u}r Hochenergiephysik, Vienna, Austria}\\*[0pt]
W.~Adam, F.~Ambrogi, T.~Bergauer, M.~Dragicevic, J.~Er\"{o}, A.~Escalante~Del~Valle, R.~Fr\"{u}hwirth\cmsAuthorMark{1}, M.~Jeitler\cmsAuthorMark{1}, N.~Krammer, L.~Lechner, D.~Liko, T.~Madlener, I.~Mikulec, F.M.~Pitters, N.~Rad, J.~Schieck\cmsAuthorMark{1}, R.~Sch\"{o}fbeck, M.~Spanring, S.~Templ, W.~Waltenberger, C.-E.~Wulz\cmsAuthorMark{1}, M.~Zarucki
% ...
\vskip\cmsinstskip
% ...
1:  Also at TU Wien, Wien, Austria\\
% ...
\end{sloppypar}
%%% END EDITABLE REGION %%%
% skeleton_end
\end{document}